\documentclass[aip,reprint]{revtex4-1}
\usepackage{graphicx}
\usepackage{color}
\usepackage{ulem}
\draft 

\begin{document}


\title{Active control of emission directionality of semiconductor microdisk lasers} 



\author{Seng Fatt Liew}
\author{Brandon Redding}
\affiliation{Department of Applied Physics, Yale University, New Haven, CT 06520, USA }
\author{Li Ge}
\affiliation{Department of Engineering Science and Physics, College of Staten Island, CUNY, Staten Island, NY 10314, USA}
\affiliation{The Graduate Center, CUNY, New York, NY 10016, USA}
\author{Glenn S. Solomon}
\affiliation{Joint Quantum Institute, NIST and University of Maryland, Gaithersburg, MD 20899, USA}
\author{Hui Cao}
\affiliation{Department of Applied Physics, Yale University, New Haven, CT 06520, USA }
\email[]{hui.cao@yale.edu}


\date{\today}

\begin{abstract}
We demonstrate lasing mode selection in nearly circular semiconductor microdisks by shaping the spatial profile of optical pump. 
Despite of strong mode overlap, adaptive pumping suppresses all lasing modes except the targeted one. 
Due to slight deformation of the cavity shape and boundary roughness, each lasing mode has distinct emission pattern. 
By selecting different mode to be the dominant lasing mode, we can switch both the lasing frequency and the output direction. 
Such tunability by external pump after the laser is fabricated enhances the functionality of semiconductor microcavity lasers. 
 
\end{abstract}


\maketitle 

Semiconductor microdisk lasers have simple geometry, small footprint, and low lasing threshold, making them excellent candidates for on-chip light sources for integrated photonics applications \cite{slusher,fujita}. 
Due to high refractive index contrast at the disk boundary, light is strongly confined by total internal reflection, forming the whispering-gallery modes (WGMs) with high-quality ($Q$) factor. 
A circular microdisk much larger than the optical wavelength support densely packed WGMs, and lasing usually occurs in multiple modes of different frequencies. 
For many applications, lasing at a single frequency is desired. 
However, it is difficult to have only one mode lasing when many WGMs of similar $Q$ exist within the gain spectrum. 
It is even more challenging to switch the lasing mode from one to another, after the laser is fabricated. \\
\indent In addition to the lack of control on the lasing frequency, the isotropic emission from circular microdisks seriously limits the application because directional output is usually required. 
One way to generate directional emission is deforming the cavity shape to break the circular symmetry \cite{stoneReview,harayamaReview,xiaoReview}. 
The deformation also enhances light leakage via evanescent tunneling and/or refraction  out of the cavity, thus reducing $Q$. 
With large deformation, refractive escape is dominant, and all the whispering-gallery-like modes have similar output directionality which is dictated by the ray dynamics \cite{schwefelJOSAB04,leePRA07, wiersigPRL08}. 
Such universality hinders switching of emission directionality even if one can select different mode to lase. 
For a weakly deformed cavity, evanescent tunneling dominates over refractive escape. 
Although the intracavity mode patterns remain nearly unaltered by slight shape deformation, the external emission can be much more sensitive \cite{lacey_PRL,creagh_1,creagh_2,ge_PRA1,ge_PRA2}. 
Even a tiny boundary variation may lead to wildly varying external fields, producing large intensity contrast between the directions of maximal and minimal emission.
The emission patterns differ significantly for the high-$Q$ modes, offering the capability of switching the output directionality by selecting different mode to lase. \\
\indent Pump engineering is an efficient way to control the lasing frequency and output direction of a semiconductor laser. 
To select a particular mode to lase, one may reduce its lasing threshold by enhancing the spatial overlap between the pump and the mode \cite{pereira,chenJOB01,chen,bisson,naidoo}. 
For instance, a ring-shaped optical pump has been used to lower the lasing threshold of WGMs in circular micropillars \cite{rex} or to produce directional emission from spiral-shaped microdisk lasers  \cite{chern}. 
The same method has been adopted for electrically pumped semiconductor lasers by patterning the electrodes to match the targeted mode profiles \cite{harayama,chern2,narimanov}. 
Switching of emission directions was realized by injecting currents to separate electrodes that had maximal overlap with individual modes \cite{FukushimaJSTQE04}. 
This technique requires \textit{a priori} knowledge of the mode profiles and demands little spatial overlap between the selected modes. 
Hence, it limits the switching capability to a few modes, and becomes practically unviable once the modes have strong spatial overlap. \\
\indent Lately active control of pump profile was demonstrated for random lasers using the spatial light modulator (SLM) \cite{lopez}. 
Even for strongly overlapped modes, adaptive shaping of the spatial profile of the optical pump enabled selection of any desired lasing frequency without prior knowledge of the mode profiles \cite{nicolas_1,nicolas_2}. 
Numerical studies also demonstrated pump-controlled directional emission from two-dimensional (2D) random lasers \cite{rotter_PRL}. \\
\indent In this letter, we apply the active control of pump profile to the semiconductor microdisk lasers.
The disks are slightly deformed due to fabrication imperfection. 
Since the $Q$ spoiling is weak, the lasing threshold remains low. 
The high-$Q$ modes, whose emission is determined by evanescent tunneling, exhibit distinct emission patterns. 
Optical gain is provided by optical pumping of the semiconductor quantum dots (QDs) embedded in the disk. 
The broad gain spectrum allows lasing in multiple high-$Q$ modes.  
These modes have strong spatial overlap near the disk boundary, which adds complexity to the mode selection process, and makes finding the best solution a nontrivial operation. 
By adaptive control of the pump profile, we are able to select different modes to be the dominant lasing mode and suppress all other lasing modes.
Consequently, the laser emission pattern is changed.    
Our results demonstrate an effective and flexible method to exploit the multimode characteristic of nearly circular microdisk lasers for switching of lasing frequency and emission direction. \\
 \begin{figure}
 \centering
 \includegraphics[scale=0.30]{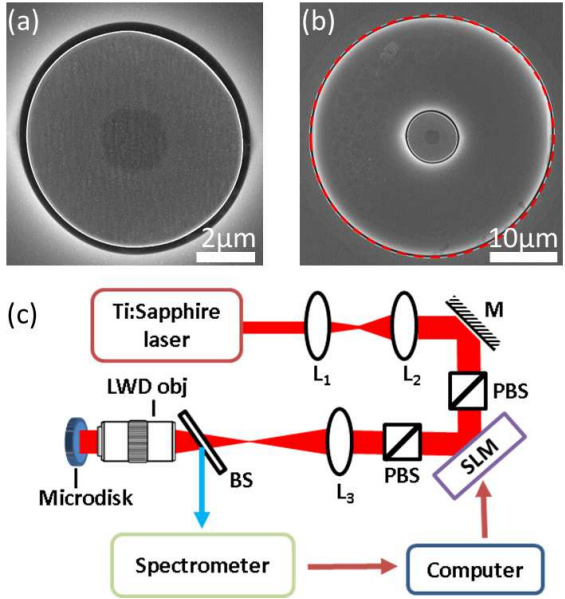}
 \caption{(a) Top-view scanning electron micrograph (SEM) of a fabricated GaAs microdisk, showing slight deformation from circular shape. Center dark area is on top of the AlGaAs pedestal. (b) Low magnification view of the microdisk in (a). Red circle denotes the outer edge of an annular air gap, which is used to measure emission pattern. (c) Schematic of optical setup: L$_1$, L$_2$, L$_3$: lens. PBS: polarizing beam splitter. SLM: spatial light modulator. BS: non-polarizing beam splitter. LWD obj: long-working distance objective lens. \label{fig1}}%
 \end{figure}  
\indent 
The sample is grown on a GaAs substrate by molecular beam epitaxy. 
It consists of a 1$\mu$m-thick Al$_{0.68}$Ga$_{0.32}$As layer and a 200nm-thick GaAs layer with three embedded layers of InAs QDs. 
Each QD layer contains 2.5 monolayers of InAs. 
Microdisks are fabricated by electron beam lithography and two steps of wet etching \cite{qinghai_PRA}. 
The first is non-selective etching of the GaAs and Al$_{0.68}$Ga$_{0.32}$As layers with HBr, forming microcylinders. 
The second step is a HF-based selective etch to undercut the Al$_{0.68}$Ga$_{0.32}$As and create a pedestal to isolate the GaAs disk from the substrate. 
Figure \ref{fig1} shows the top-view scanning electron micrograph (SEM) of a fabricated disk. 
The disk at the center is surrounded by an annular air gap that separates the disk from the unetched GaAs layer. 
The outer circle (red dashed line), with a radius of 18$\mu$m, is used to measure the emission pattern. 
The non-selective etching process is not perfectly isotropic, thus the disk shape deviates from the original design of a circle, as can be seen in Fig. \ref{fig1}(b). 
In addition, the etching also creates surface roughness at the disk boundary. 
From the high-resolution SEM, we extract the disk boundary and fit it in the polar coordinates as 
 $ \rho(\theta) = R\left[1 + a\cos(2 \theta+\alpha) + b\cos(3 \theta+\beta) \right] $,
where $R = 3.71\mu$m, $a = 0.024$, $b=0.0089$, and $\alpha = 1.38$, $\beta = 0.13$. 
The dominant deformation originates from the $\cos(2\theta)$ modulation, but the contribution from the $\cos(3\theta)$ is non-negligible. 
Note that both modulations have very small magnitudes, $a,b \ll 1$, confirming the cavity is nearly circular and the output is dominated by evanescent tunneling.\\
\begin{figure}
\centering
\includegraphics[scale=0.32]{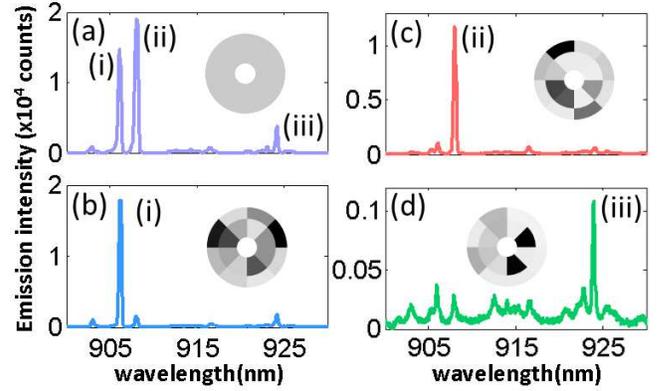}
\caption{Emission spectra of the microdisk shown in Fig. 1(a) when pumped with different spatial patterns. The total pump power is kept at 2.2mW. Three modes lase with a homogeneous ring pump (a). Each of the three modes, (i-iii), becomes the dominant lasing mode (b-d) after the optimal pump profile is found by the genetic algorithm. The inset in each panel is the final pump pattern. Darker color corresponds to higher pump intensity. \label{fig2}}
\end{figure}
\indent The fabricated samples are tested in a liquid helium cryostat at temperature $\sim$ 10K. 
Optical excitation is provided by a mode-locked Ti:Sapphire laser ($\lambda$ = 790nm) operating at 76MHz with 200fs pulses. 
As shown schematically in Fig. \ref{fig1}(c), the pump beam is first expanded by a telescope to cover the entire surface of the SLM (Hamamatsu X10468-02). 
The SLM is positioned between a pair of polarizers to introduce intensity modulation on the pump beam. 
The spatially modulated pump pattern is then projected onto the top surface of a microdisk. 
Since the AlGaAs pedestal causes light leakage from the GaAs disk to the substrate, the high-$Q$ modes avoid the central region of the disk. 
Thus the pump region is set as a ring. 
It is divided into two subrings, each is further divided into eight sections in the azimuthal direction. 
Optical power within each section can be modulated separately using the SLM while the total power is kept constant. \\
\indent In order to select one mode to lase while suppressing all other modes, we adopt the genetic algorithm in MATLAB to search for the optimum spatial pump profile. 
The cost function is defined as $G = I_m/I_o$, where $I_m$ is the intensity of the targeted mode, and $I_o$ is the highest intensity among all other modes in the spectra. 
The algorithm starts with an initial population of ten pump patterns where one of them is a homogeneous ring  and the rest are random. 
The emission spectrum for each pump pattern is recorded, and the cost function is evaluated. 
The pump patterns are ranked by their cost function, and the ones with higher ranking have a larger chance of being chosen as ``parents'' to generate ten ``children''. 
The first two are the pump patterns with the highest ranking among the parents, which are copied to the next generation without changes. 
The next six childrens are generated through crossover by randomly selecting different parts of the pump patterns from a pair of parents and combine them. 
The last two are created through mutation by making random changes to a single parent. 
The mutation rate decreases with generation and is set to zero after ten generations. 
After ten generations, we reset the initial population to include the highest ranked pump pattern from the previous trial and nine randomly generated patterns, and repeat the genetic algorithm to search unexplored parameter space for a better solution. 
During the optimization process, the total pump power is kept constant, the pump energy is merely re-distributed to different regions of the disk. 
Between successive pump patterns test, the microdisk is pumped with a homogeneous ring to eliminate any residual thermal effect from the previous pump profile. \\
\begin{figure}
\centering
\includegraphics[scale=0.28]{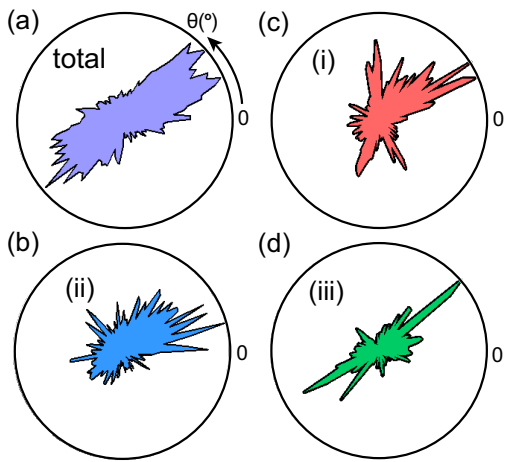}
\caption{Angular distribution of emission intensity measured for the microdisk in Fig. 1(a). The corresponding emission spectra are shown in Fig. \ref{fig2}(a-d). 
The total emission from all three lasing modes is bi-directional (a), but each of them, (i-iii), exhibits distinct emission pattern (b-d). The mode labeling is the same as that in Fig. \ref{fig2}.  \label{fig3}}%
\end{figure}
\indent Figure \ref{fig2} presents the results for adaptive pumping of the microdisk shown in Fig. 1. 
When the pump intensity is uniform across the annular pump region, the emission spectrum contains three major peaks at the pump power of 2.2 mW [Fig. \ref{fig2}(a)].
No additional lasing peak is found beyond the spectral range of Fig. \ref{fig2}(a). 
We are able to make any one of the three to be the dominant lasing mode after optimizing the pump pattern in 30 generations. 
Figure \ref{fig2}(b-d) are the emission spectra after optimizing modes (i)-(iii) respectively, and the insets are the optimized pump patterns. 
The emission intensity of the selected mode changes slightly but the intensities of non-selected modes are greatly reduced.
For example, when mode (iii) is chosen to be the dominant lasing mode [Fig. \ref{fig2}(d)], the intensities of mode (i) and (ii) are an order of magnitude lower than under the homogeneous pumping [Fig. \ref{fig2}(a)]. \\
\indent Next, we investigate the output directionality of the lasing modes. 
The in-plane emission from the disk boundary propagates to the outer edge of the air gap [red circle in Fig. \ref{fig1}(a)] and is scattered out of the plane. 
The scattered light is imaged from the top of the disk by an objective lens onto a CCD camera, and its intensity distribution reflects the emission pattern. 
When three modes lase simultaneously with homogeneous pumping [Fig. \ref{fig2}(a)], the total laser emission is bidirectional as shown in Fig. \ref{fig3}(a). 
It is attributed to the dominant $\cos(2 \theta)$ modulation of the cavity boundary and light is emitted tangentially from the two places of the highest curvature on the cavity boundary. 
Next we place a narrow-bandpass filter in front of the camera to select a single lasing mode. 
Figure \ref{fig3}(b-d) show the angular distribution of emission intensity for each of the three lasing modes. 
It is evident that the three modes have distinct emission patterns. 
When one of them becomes the dominant lasing mode by selective pumping, we remeasure its emission pattern and find it remains the same. 
Hence, the modes themselves are barely modified by the redistribution of pump energy. 
This result differs from the weakly scattering random lasers \cite{nicolas_1,nicolas_2}, because in our case the modes are strongly confined due to high index contrast at the disk boundary. \\
\indent To understand the modal dependent output directionality, we extract the cavity shape from the SEM and perform numerical simulation. 
The lasing modes usually correspond to the high-$Q$ modes in the passive cavity, whose frequencies are within the gain spectrum. 
Thus we numerically calculate the high-$Q$ modes in the passive disk using the finite element method (COMSOL). 
Since the disk radius is much larger than the disk thickness, a microdisk can be treated as a 2D cavity with an effective index of refraction $n=3.13$ \cite{qinghai_PRA}. 
To simulate the open boundary condition, the disk is placed in air, which is surrounded by a perfectly matched layer (PML) to absorb all outgoing waves. 
Since we do not know the exact temperature of the GaAs disks when optically pumped inside the liquid-Helium cryostat, we cannot get the accurate value of the refractive index $n$ to match the numerically calculated modes with the experimentally measured lasing peaks.
Instead of a quantitative comparison, our numerical simulation aims to provide a qualitative understanding of the characteristic of the lasing modes. \\
\begin{figure}
\centering
\includegraphics[scale=0.26]{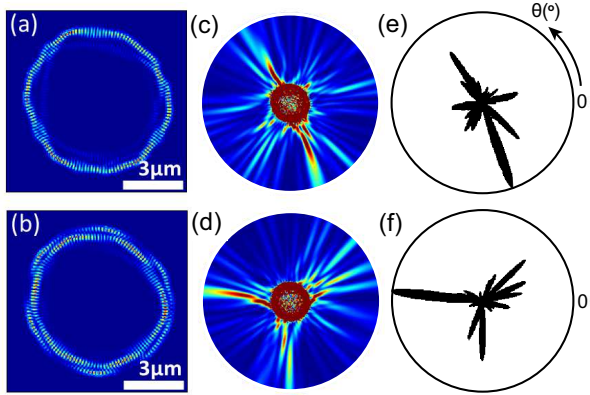}
\caption{Numerical simulation of two high-$Q$ modes, one at $\lambda$ = 913.4nm (a,c,e), the other at $\lambda$ = 927.7nm (b,d,f), in the microdisk shown in Fig. 1(a).   
(a,b) Spatial distribution of electric field intensity, resembling the WGM but distorted due to boundary roughness. 
(c,d) Electric field intensity distribution outside the disk, highlighted by saturating the intensity inside the disk, showing the two modes have distinct outputs. 
(e,f) Angular distribution of emission intensity, convolved with the experimental resolution, at a distance of 18$\mu$m from the disk center. \label{fig4}}%
\end{figure}
\indent Figure \ref{fig4} presents the results of two high-$Q$ modes in the wavelength range of InAs QD gain spectrum.  
Their quality factors $Q = \omega_r/2\omega_i$ are 67000 and 69000 respectively, where $\omega_r-i\omega_i$ is the complex frequency of the cavity mode, and the imaginary part $\omega_i$ is inversely proportional to the mode lifetime in the open cavity. 
As shown in Fig. \ref{fig4}(a,b), both modes are spatially localized near the cavity boundary, similar to the WGMs. 
However, their intensities are not uniformly distributed in the azimuthal direction, because of boundary roughness. 
Outside the cavity, the azimuthal variation of the field intensity becomes much stronger as seen in Fig. \ref{fig4}(c,d). 
The two modes exhibit distinct intensity distributions outside the cavity, indicating that light output via evanescent tunneling is dramatically different. 
Figure \ref{fig4}(e) and (f) are the angular distributions of emission intensity at a distance of 18$\mu m$  from disk center, similar to the experimental measurement. 
The calculated intensity distribution is convolved with the experimental resolution of $\sim 0.9$ $\mu$m.
We have simulated other high-$Q$ modes and observed similar phenomena. 
Finally, we checked that without boundary roughness, the microdisk defined by the fitted boundary curve $\rho(\theta)$, contains only high-$Q$ modes with similar emission patterns. 
This result confirms that boundary roughness is responsible for the distinct emission directionality of individual high-$Q$ modes. 
The sidewall roughness not only  modifies the local evanescent tunneling, but also induces multimode coupling which affects the far-field pattern \cite{ge_PRA2}. 
The diversity in the output directionality among the high-$Q$ modes enables the switching of emission direction by selecting different modes to lase with adaptive pumping. \\
\indent In conclusion, we demonstrate selection of lasing modes with directional emission in weakly deformed semiconductor microdisks by adaptive pumping. 
Despite strong spatial overlap of the lasing modes, we are able to select any one of them to be the dominant lasing mode by suppressing all other modes. 
Slight shape deformation and sidewall roughness due to fabrication imperfection creates directional emissions that are mode dependent. 
Combining these features, both lasing frequency and emission pattern can be switched by external pump, after the laser is fabricated.  
This method may be extended to electrically pumped microdisks by using multiple eletrodes to modulate the spatial profile of current injection. 

We thank Jan Wiersig, Nicolas Bachelard, Patrick Sebbah, Douglas Stone, Sebastien Popoff and Alex Cerjan for useful discussions. This work is supported partly by NSF under Grant Nos. ECCS-1128542 and DMR-1205307.  Facilities use is supported by YINQE and NSF MRSEC Grant No. DMR-1119826.


\end{document}